# High energy Coulomb-scattered electrons for relativistic particle beam diagnostics

P. Thieberger, Z. Altinbas, C. Carlson, C. Chasman, M. Costanzo, C. Degen, K. A. Drees, W. Fischer,
D. Gassner, X. Gu, K. Hamdi, J. Hock, A. Marusic, T. Miller, M. Minty, C. Montag, Y. Luo and A.I. Pikin,

Brookhaven National Laboratory, Upton, NY 11973, USA

S.M. White

European Synchrotron Radiation Facility, BP 220, 38043 Grenoble Cedex, France.

A new system used for monitoring energetic Coulomb-scattered electrons as the main diagnostic for accurately aligning the electron and ion beams in the new Relativistic Heavy Ion Collider (RHIC) electron lenses is described in detail. The theory of electron scattering from relativistic ions is developed and applied to the design and implementation of the system used to achieve and maintain the alignment. Commissioning with gold and $^3$He beams is then described as well as the successful utilization of the new system during the 2015 RHIC polarized proton run. Systematic errors of the new method are then estimated. Finally, some possible future applications of Coulomb-scattered electrons for beam diagnostics are briefly discussed.

## I. INTRODUCTION

Instrumentation providing accurate information on particle beam properties and behavior in accelerators is essential for their operation. The challenge of performing reliable and often delicate measurements in the harsh particle accelerator environment provides strong incentives for exploring new approaches. We describe a new type of beam diagnostic device for high energy particle accelerators based on the Coulomb scattering of electrons by the beam particles. Measuring the deflection of low energy electron beams by the macroscopic fields generated by the high energy particle beam to be characterized, the so-called electron wire, was proposed in the late 1980s and early 1990s [1, 2, 3, 4] and later implemented in several systems, including the use of electron ribbons instead of the pencil beams [5, 6, 7, 8]. The system we describe here is a new noninvasive beam diagnostic tool also based on the Coulomb interaction of low energy electrons with relativistic particle beams, but in this case the interaction is due to small impact parameter collisions of a small fraction of the electrons with individual beam particles leading to large momentum transfers. This mechanism is the so-called Rutherford scattering, named after Ernest Rutherford who, in 1911 [9], discovered the atomic nucleus by studying the scattering of alpha and beta particles from stationary targets. Our targets, the ion beam particles, far from being stationary, are moving at relativistic velocities. The theory describing the interaction is the same in the frame of reference co-moving with the particle beam. Using this theory, we can predict the energies and angular distribution of the scattered electrons by coordinate transformation to the laboratory frame. In the laboratory frame, some of these electrons acquire energies up to several MeV, making them easy to detect even after traversing thin vacuum windows, thus allowing the use of simple scintillation detectors in air.

Based on these ideas, we developed a non-invasive diagnostic tool called electron beam backscattering detector (eBSD) [10] to accurately align the electron and proton beams in the new Brookhaven National Laboratory (BNL) electron lenses for the partial compensation of the head-on beam-beam effects that limit the luminosity [11].

In the following sections we review the theory of electrons scattered form relativistic ions, we then describe the principle of applying this phenomenon to the alignment of electron and ion beams in the RHIC electron lenses and we describe the implementation of the backscattered electron detectors. We then report on the commissioning of the systems with gold and helium beams and, finally, we describe in detail the successful utilization of this new diagnostic instrument during the 2015 RHIC polarized proton-proton operations run (henceforth: "run"). Based on the experience and data from these runs, we then provide an analysis and estimates of systematic errors of this alignment method.

During the first commissioning run of these systems [10, 12] it was discovered that energetic scattered electrons are also generated by the interaction of the relativistic particles with the electrons of residual gas atoms. We mention here the possible use of these electrons in other non-invasive beam diagnostic instruments not requiring a low energy electron beam, and we suggest the possibility of using instruments similar to the eBSD for the alignment of other configurations involving ion and electron beams such as hollow beams for collimation or for halo monitoring, and long range beam-beam compensators [13].

Finally, we suggest that detecting the electrons scattered from an "electron wire" may be an alternative, and more intensity-independent way to obtain profiles as compared to the present measurements of small deflections caused by the macroscopic fields generated by the beam. Time resolved measurements of this type, in addition to providing transverse beam profiles, could also provide longitudinal bunch profiles and diagnostics for "head-tail" perturbations.

## II. THEORY OF ELECTRONS SCATTERED FROM MOVING TARGETS

To first order in the fine structure constant, the Coulomb scattering of relativistic electrons by nuclei is described by the Mott formula which in the rest frame of the nucleus is written [14]:

$$\frac{d\sigma}{d\Omega} = \frac{Z^2}{4}\left(\frac{e^2}{E}\right)^2 \frac{1}{sin^4(\theta/2)} \times \left[1 - \left(\frac{pc}{E}\right)^2 sin^2\frac{\theta}{2}\right] \times \left[1 + \frac{2E\, sin^2(\theta/2)}{M_A c^2}\right]^{-1} \qquad 1]$$

where $\Omega$ is the solid angle, $\sigma$ the cross section, $Z$ the atomic number, $M_A$ the mass of the nucleus, $e$ the elementary charge, $E$ and $p$ the energy and momentum of the electron in the frame of the nucleus and $\theta$ the electron scattering angle in that frame. The first term is the classical Rutherford cross section and the two bracketed terms come from the 1/2 spin of the electron and the nuclear recoil respectively. A small correction for the nuclear magnetic moment has been neglected. Another correction that has been neglected is the one for Bremsstrahlung, which can be significant. A more complete theoretical treatment will probably be required to make good quantitative predictions, especially for even higher energy protons or ions.

Values of this cross section are computed at small angular intervals and then relativistic transformations to the laboratory frame of the cross sections, the angles and the energies lead to results such as plotted in Fig. 1. Such plots are useful for rough estimates of counting rates, but detailed comparisons are difficult due to the complicated nature of the spiraling electron trajectories in our particular application (see next section).

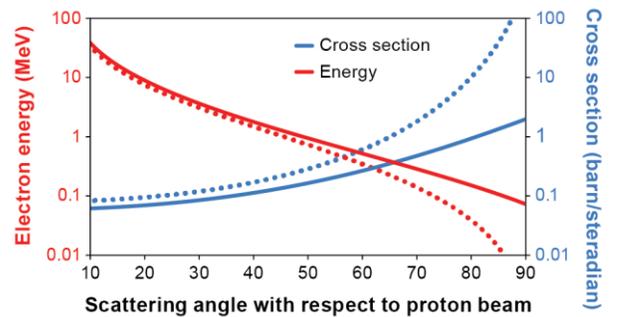

FIG. 1. The solid lines show calculated energies and approximate scattering cross sections for 5 keV electrons backscattered by 250 GeV protons. The dotted lines correspond to the same quantities, but for 10 eV electrons as a qualitative indication of energetic electrons generated by the interaction of the beam with the residual gas and/or with low energy electrons captured in the potential well of the beam.

For the same example, the laboratory angle is plotted in Fig. 2 as function of the angle in the proton frame of reference, both angles being shown in this case with respect to the original electron propagation direction. We see that this is a rather extreme example of relativistic beaming, also referred to as the "headlight effect". Electrons scattered forward in the proton frame at angles larger than ~0.1 degrees appear in the laboratory at angles larger than 90 degrees, i.e. they are backscattered.

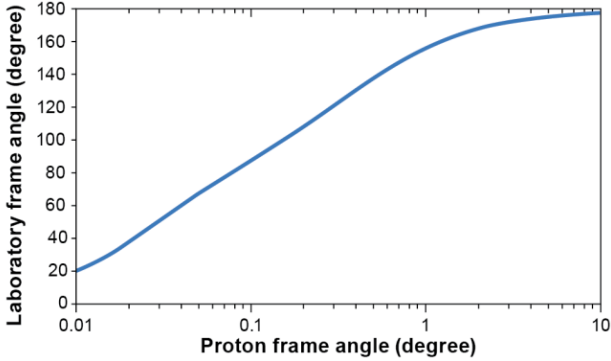

FIG. 2 Angles with respect to the initial electron beam direction for 5 keV electrons scattered of 250 GeV protons moving in the opposite direction. For this example, the energy of the electrons in the proton rest frame is 157 MeV.

## III. RHIC ELECTRON LENS BEAM ALIGNMENT

The partial compensation of the head-on beam-beam effect in RHIC is necessary for mitigating the limit imposed by this effect on the achievable proton-proton beam luminosities [11]. Electron lenses (e-lenses) [15, 16, 17, 18] consisting of low energy (in our case ~5 keV), high intensity (~1 A), magnetized electron beams [19], can provide the precise non-linear focusing properties necessary to effect such compensations. A schematic view of the two electron lenses that are being used for this purpose in RHIC is shown in Fig. 3.

The precise alignment overlap of the electron and ion beams is an important prerequisite for achieving maximum compensation and for avoiding deleterious effects on the proton beam emittance [20, 21, 22]. Over the 2.4 m interaction region in the up to 6 T solenoid, the centers of these, as small as 300 μm rms wide beams, need to be separated by less than 50 μm. The precision achievable with the installed beam position monitoring system is not sufficient for ensuring this result, especially in view of electronic offsets that are not identical for electrons and ions [23, 24]. Besides, in order to generate Beam Position Monitor (BPM) signals, the dc electron beam needs to be modulated and this modulation may affect electron beam stability during operation. As mentioned above, electrons in the electron beam, backscattered by the relativistic protons can provide a signal proportional to the electron-proton luminosity that can be used to maximize the overlap of the two beams [10].

The lensing effect of an e-lens on the relativistic protons is due to the macroscopic electric and magnetic fields produced by the Gaussian-shaped electron beam. In other words, it is the collective long-range Coulomb interaction of the electrons with individual ions that affects the trajectory of these ions. The vast majority of the electron trajectories are only slightly affected transversely [25] since their trajectories are confined by a strong solenoid magnetic field. There is, however, a finite probability for ion-electron collisions with impact parameters that are so small as to produce a significant electron scattering angle, imparting at the same time considerable momentum and energy to the scattered electrons. Large scattering angles, correlated with high energies, result in energetic electrons spiraling backwards (towards the electron gun) along the magnetic field lines. Some of these backscattered electrons are intercepted and counted by a scintillation detector placed in air, behind a thin vacuum window.

Figure 4 shows the simulated projected trajectories of two electrons backscattered by 250 GeV protons in a 6 T solenoid at angles of ±50º, one upwards and the other one downwards. As the electrons spiral towards the detector, the radii of their trajectories grow as they encounter lower fields. The upward drift of the trajectory envelopes of the back-scattered electrons is due to the horizontal bend in the field [26]. The higher energy of the scattered electrons as compared to the electrons in the primary beam makes this drift appreciable for the latter, while it is negligible for the former. This upward drift of the scattered electrons is helpful in separating the electron trajectories from the primary electron beam, thus facilitating their detection.

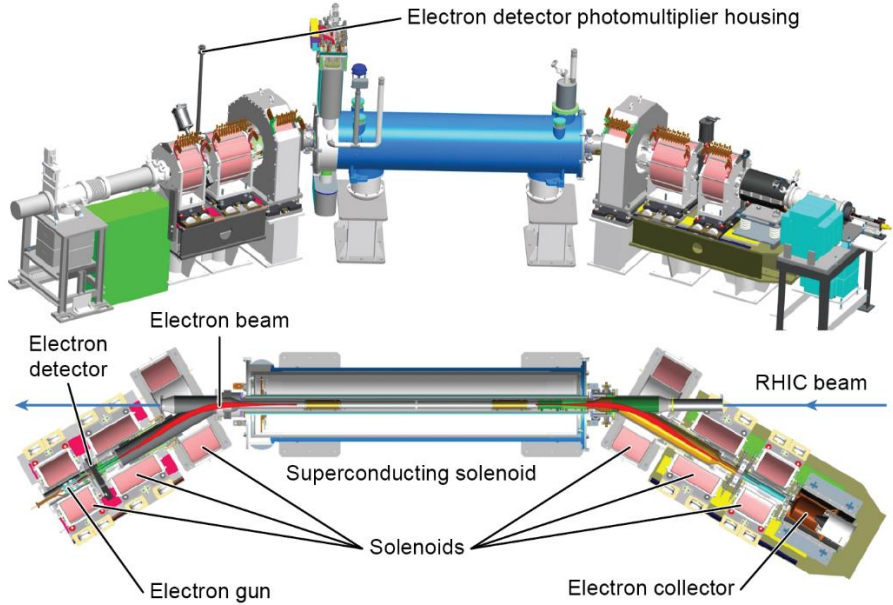

FIG. 3. Perspective and top views of one of the RHIC electron lenses [11, 12], showing the backscattered electron detector location close to the electron gun. In reality, the light guide and photomultiplier tube are enclosed in a heavy and light-tight magnetic shield to protect the PMT from the stray fields of the nearby magnets.

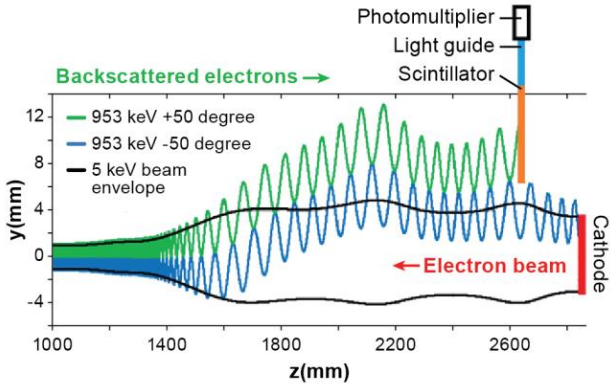

FIG. 4. Computer simulated trajectories (blue and green curves) of two scattered electrons generated inside of the 6 T solenoid (see text). Only the first 200 mm of this 2400 mm long superconducting solenoid is included at the left (from Z=1000 mm to Z=1200 mm). Three weaker solenoids (not shown) guide the electron beam from the cathode towards the 6 T region (see Fig. 3)

Each RHIC electron lens is equipped with an eBSD consisting of a small plastic scintillator ($7.4 \times 7.9 \times 20.6$ mm$^3$) attached to a 1.2 m long light guide leading to a small magnetically shielded photomultiplier (PMT) tube (Hamamatsu R3998-02) [27]. The signals from this PMT reach the instrumentation rack through a ~90 m long, 50 Ω coaxial cable and are amplified and connected to a fast discriminator, the output pulses of which are used to determine the counting rates. The long light guide is necessary to keep the PMT far enough from the adjacent magnets so as to enable adequate shielding. This scintillation detector assembly is mounted in air in a vertical shaft, at the bottom of which there is a 0.1 mm thick titanium alloy vacuum window facing the scintillator. The vertical position of the detector shaft can be selected so as to locate the bottom of the scintillator at any position from 1 mm to 25.4 mm from the edge of the primary electron beam. This position adjustment can be used as an intensity range selector. An insulated tungsten block ($35 \times 4.9 \times 7.6$ mm$^3$) with current detection provides some protection against electron beam heating, should the position interlock and limit switch fail. Such a failure actually occurred during commissioning and a RHIC vacuum failure was avoided thanks to the tungsten block even though indirect heating was sufficient to melt the scintillator. Since the scintillation detector was in air, it was easily replaced.

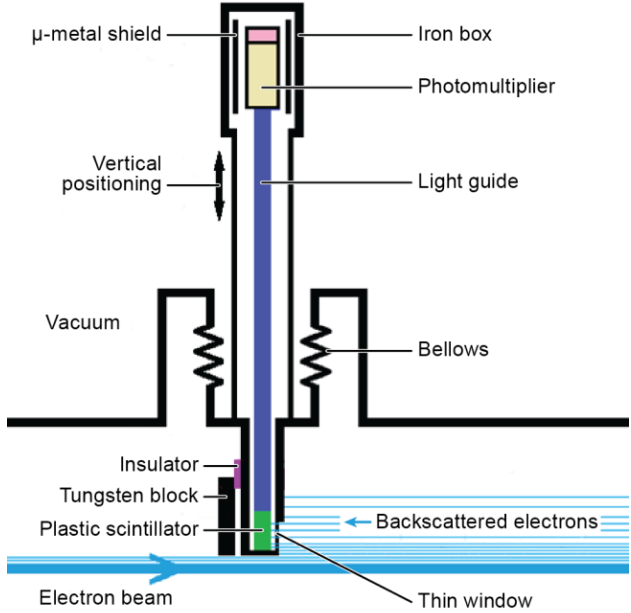

FIG. 5. Not-to-scale schematic of the eBSD scintillation detector and its housing. Electrons backscattered by the relativistic ion beam reach the plastic scintillator after traversing a thin titanium alloy vacuum window. The light from the scintillator is converted to electrical signals by means of a well shielded photomultiplier tube. A tungsten block protects the back of the detector cavity from accidental heating by the electron beam.

A not-to-scale schematic and a cutaway drawing of the detector housing are shown in Figs. 5 and 6 respectively.

The design and fabrication of the 100 μm thick Ti-6Al-4V alloy window was a critical aspect of this project since the electron energy loss had to be minimized while guaranteeing the integrity of the RHIC vacuum system. Fig.7 shows the stopping power [28], and the calculated energy loss in a 0.1 mm thick titanium alloy window.

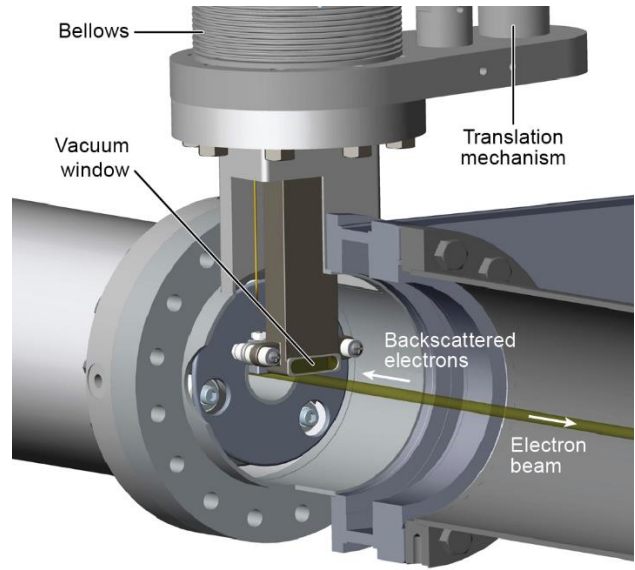

FIG. 6. Cutaway drawing of the detector housing, the vertical translation mechanism and the location of the 0.1 mm thick titanium alloy vacuum window in its lowest position. The area of this window is $25.4 \times 6.35$ mm$^2$. The vertical position can be changed by up to 25.4 mm away from the electron beam. That highest position was the one used for most measurements to minimize excessive counting rate issues.

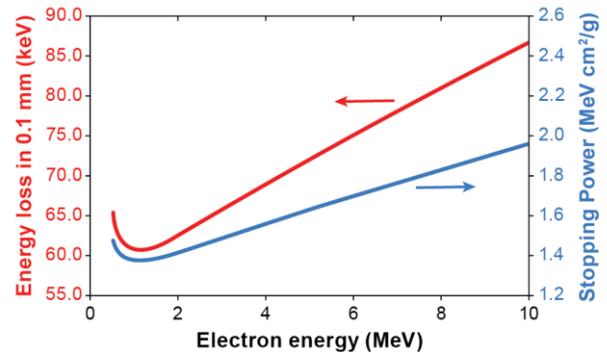

FIG. 7. Stopping power (blue) [28] and calculated energy loss (red) in titanium as function of electron energy.

We see that the energy loss in a 0.1 mm thick titanium window is acceptable for electrons of a few hundred keV and up. The design and dimensions of the window and detector housing are shown in Fig. 8 and the corresponding stress analysis is presented in Fig. 9.

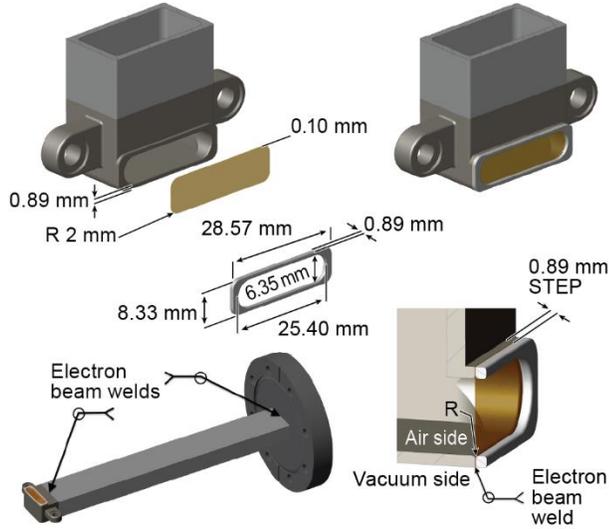

FIG. 8. Design and fabrication details of the cavity and the 0.1 mm thick titanium alloy window. The light guide is inserted in the cavity with the plastic scintillator facing the thin window. The electrons traverse this window from the vacuum outside of the cavity to the inside that is at atmospheric pressure.

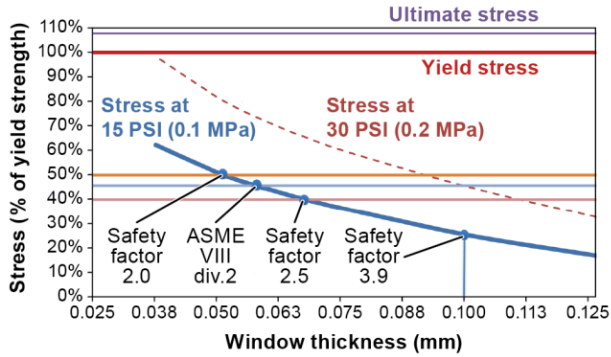

FIG. 9 Finite element stress analysis results for the 25.4 × 6.9 × 0.1 mm$^3$ Ti-6-Al-4V alloy window. There is a safety factor of 3.9 at atmospheric pressure. The window was pressure tested up to three atmospheres without bursting.

The detector cavity and the 0.1 mm thick window shown in Fig. 8 were fabricated using Ti-6Al-4V alloy which provided the desired strength and relatively small electron energy loss (see Fig. 7). The stress analysis shown in Fig. 9 as well as pressure tests proved that the safety factor is larger than 3.

## IV. COMMISSIONING WITH GOLD AND $^3$He BEAMS IN RHIC

The commissioning of the eBSDs was started during the 2014 100 GeV/nucleon gold-gold run. The first proof-of-principle horizontal and vertical beam separation scans are shown in Figs. 10 and 11.

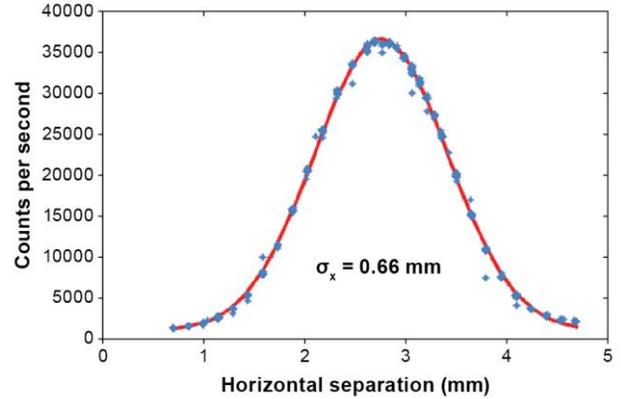

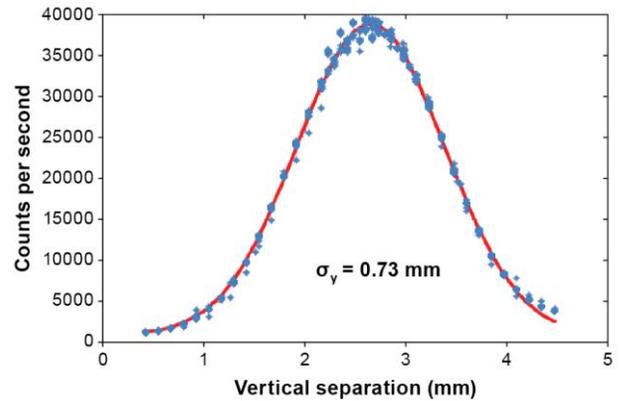

FIG. 10 Horizontal and vertical beam separation scan obtained by steering the 5 keV electron beam with respect to the 100 GeV/nucleon gold beam.

The measured widths are both about 25% larger than the sums in quadrature of the gold and electron beam widths. This small discrepancy could be due to residual angular misalignments, to small ion beam motions or to errors in the gold beam beta function estimates.

Soon after obtaining these results, a beam alignment optimization system was implemented based on automatically maximizing both the eBSD counting rates as function of horizontal and vertical positions and angles. This system is based on a program (LISA) [29] that was developed many years ago and used since then to maximize the ion-ion luminosities for the RHIC experiments by maximizing the coincidence rates from the zero degree calorimeters (ZDCs) [30].

After the gold run was completed, there was a brief opportunity for commissioning the eBSD system with a $^3$He beam. This was important since gold scattering cross sections are much larger than the cross sections for protons, and therefore the gold beam tests were not representative of the situation with protons. The cross section for $^3$He is ~4 times larger than for protons (see eq. 1) but the intensity was smaller. The counting rates for $^3$He were similar to the ones expected for protons. Horizontal and vertical separation scans are shown in Fig. 11. These data are from a manual LISA eBSD scan obtained by displacing the gold beam by means of a set of steering correctors forming closed horizontal and vertical bumps.

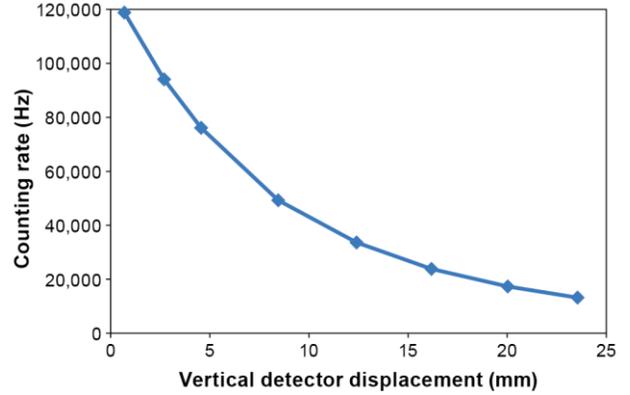

FIG. 12. Counting rate as function of the vertical displacement of the eBSD detector. These results were obtained with a 100 GeV/nucleon $^3$He beam consisting of 93 bunches with $4.7\times10^{10}$ ions per bunch and a 6 keV, 88 mA electron beam. The zero of the displacement scale corresponds approximately to the center of the detector being 5 mm above the center of the electron beam. The operating point was the rightmost point on the chart, at 24 mm. The counting-rate slope at that point is approximately 6.8 %/mm.

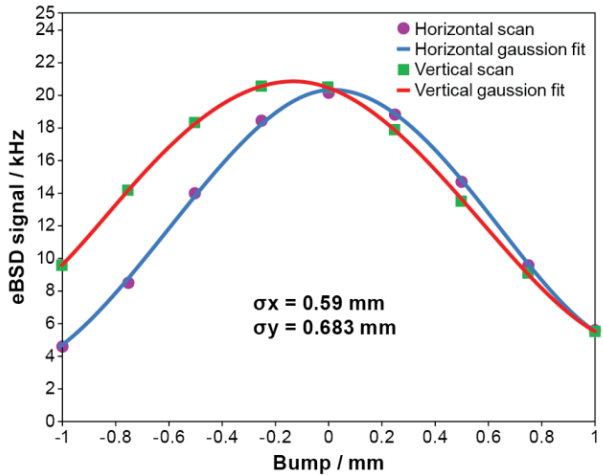

FIG. 11 Manual beam separation scans obtained by stepwise steering the $^3$He beam with closed bumps utilizing part of the algorithm developed for the automated alignment optimization system based on the LISA program [29].

During the $^3$He run the vertical positioning mechanism was utilized for the first time since the counting rates with gold had always been so large that the fully retracted position had to be used. Figure 12 shows the counting rate as a function of detector position for a 100GeV/nucleon $^3$He beam consisting of 93 bunches with $4.7\times10^{10}$ ions per bunch and a 6 keV, 88 mA electron beam

## V. UTILIZATION OF THE eBSDs DURING THE 2015 RHIC POLARIZED PROTON RUN

During this two-month run, the eBSDs were used routinely as the main alignment and monitoring tools for the electron and proton beam overlap in both electron lenses, without any system failures.

To ensure optimal pulse height discriminator settings, rejecting low amplitude noise while minimizing any impact of gain shifts, a pulse-height analysis system was implemented shown schematically in Fig. 13.

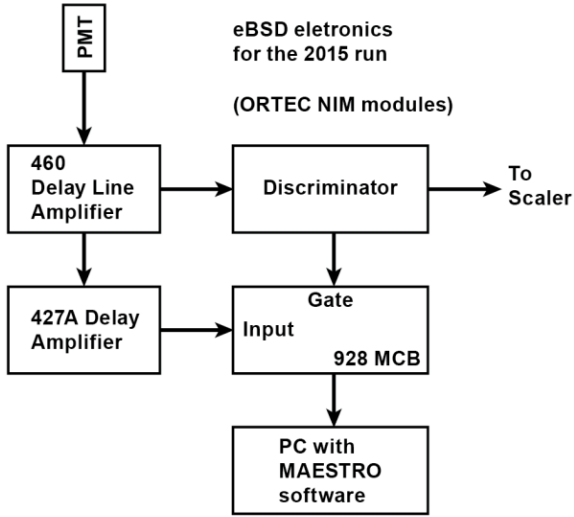

FIG. 13 Schematic of the pulse height analysis system used to optimize the discriminator setting. The Multichannel Analyzer (MCA) used here consists of an Ortec Multichannel Buffer (MCB) connected to a computer running the Maestro analysis and control software [32].

Figure 14 shows pulse height spectra and selected discriminator settings. The pulse-height resolution is poor, mainly due to the small photon collection efficiency through the thin, 1.2 m long light guide. However, excellent signal-to-noise ratios are achieved by adequate selection of the discriminator setting.

The stability of the system was such that only one slight readjustment was performed during the entire period. By the end of the run, there was a ~12% pulse height reduction measured with a precise light pulse generator [31]. This slight pulse height reduction is illustrated in Fig. 15 which shows screen shots of counting rates as a function of pulse height from the pulse height analyzer program, Maestro [32]. All settings were identical and the pulse height of the light pulser peak is reduced by about 12% after two months of continuous use. This reduction may indicate slight radiation damage of the 1.2 m long light pipe and/or of the fiber carrying the light to its far end. It would not reflect any reduction in the scintillation efficiency. Rather than using this light pulser as a reference, it would have been better to install a very weak radioactive source for continuous, end-to-end gain verification. Modest reductions in pulse height can be easily compensated by adjusting the PMT high voltage. If necessary, the detector assembly can be easily replaced.

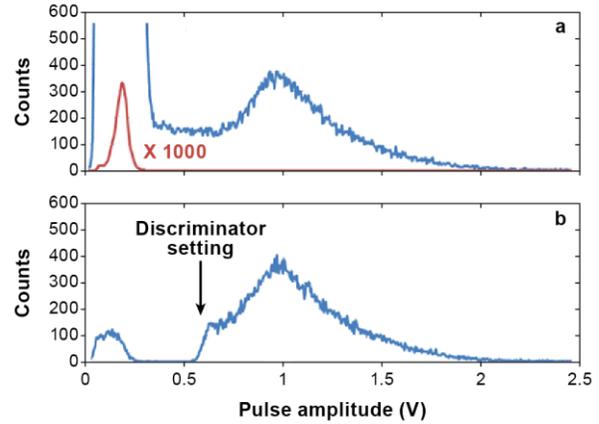

FIG. 14 Pulse height spectra from the scintillation counter used in one of the RHIC electron lenses. In the upper spectrum all pulses are accepted, while in the lower one, pulse amplitudes below the discriminator setting are rejected. The discriminator effectively suppresses the high intensity low amplitude noise (see the red curve, which is the original histogram plotted with a scale change of a factor 1000) . The small remaining peak to the left in (b) may be an artifact of the pulse height analyzer gating system.

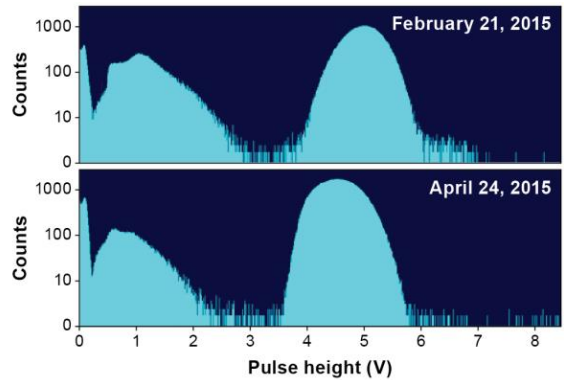

FIG.15 . Pulse height spectra screen shots obtained with the pulse-height analyzer software [32]. The logarithm of counting rates are displayed as a function of pulse heights. The peaks to the right were obtained using identical settings of a precision light pulse generator. The slight shift of this peak indicates a ~12% light transmission loss due perhaps to slight radiation damage of the 1.2 m long light guide or of the fiber carrying the light signal to the far end of the light guide.

The optimization of the beam alignment was largely automated by using the LISA algorithm described above [29]. To simplify the angular adjustments, the algorithm was modified, so as to rotate the beams around the centers of the respective lenses rather than around the proton-proton crossing point which is 3.3 m away from the center of the lenses. No interactions occur at this crossing point because the two beams are at different heights being separated by 10 mm or approximately 20 sigmas.

During this run, compensation with close to maximum electron current was only used at the beginning of each store and was then reduced in steps as shown in Fig. 16. This optimized compromise provided the best integrated luminosity by utilizing the electron lenses when most necessary for compensation, while minimizing their impact on beam lifetime and emittance [11].

During this period, a system that sorts the eBSD signals according to their arrival time was tested [10]. For this purpose, time digitizers [33] were started by the eBSD signals and stopped by a signal synchronized to the RHIC revolution frequency. While this system wasn't utilized, we mention it here because it may be used in the future and may also be of interest for other applications (see next section).

Figure 17 shows a time spectrum obtained with signals from the proton beam scattering electrons from the residual gas. Variations in count rate from peak to peak are consistent with variations in the individual bunch intensities. The 111-bunch structure representing one turn ($T_{rev}$ = 12.8 μs) is shown at the top, and the last 10 bunches followed by part of the abort gap is shown at the bottom. The fast rising part of the peaks represents electrons arriving early which tend to originate closer to the gun, while the right side of the peaks, with a gentler slope, corresponds to later electrons from the other end of the interaction region. This effect could, in principle, be used as an aid for angular tuning of overlapped electron and proton beams.

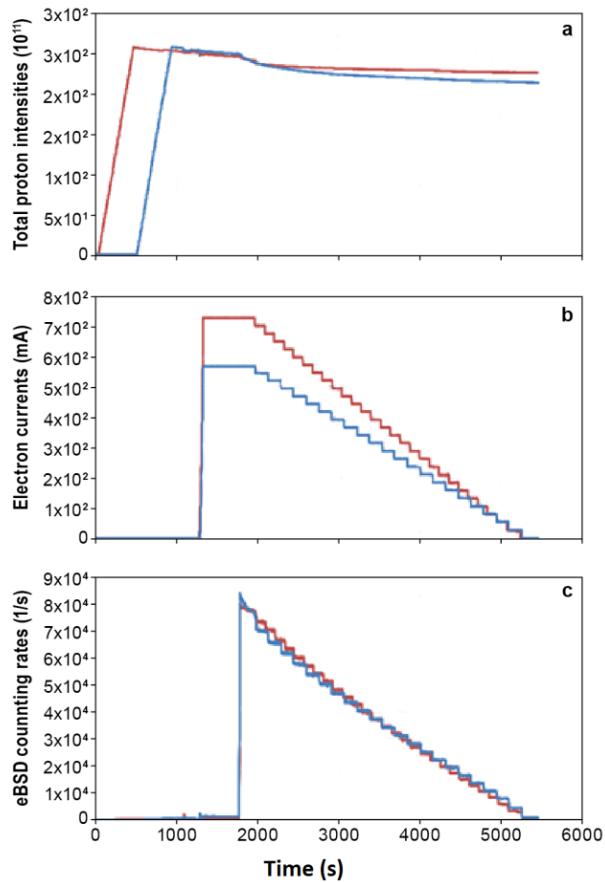

FIG.16. The total proton intensities (a), RHIC electron lens electron current intensities (b), and corresponding eBSD counting rates (c) are shown as a function of time. The electron beam intensities are reduced in steps to optimize the benefits of the RHIC electron lenses [11]. The two colors indicate the results from each of the two RHIC rings.

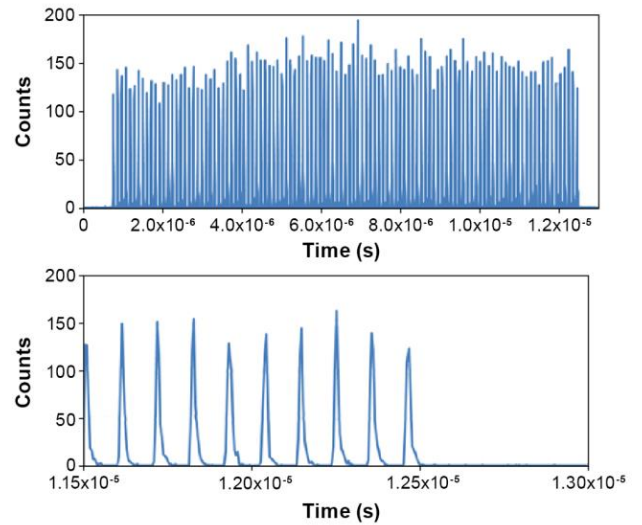

FIG.17. Time-of-flight spectra of electrons generated by the interaction of the relativistic proton beam with the residual gas. Data were accumulated for 5 minutes due to the low counting rates. The top chart shows the RHIC 111-bunch pattern and portions of the abort gaps preceding and following that pattern. The bottom chart is an expanded view of the last 10 bunches to show the asymmetric shape of the peaks (see text).

## VI. SYSTEMATIC ERROR ESTIMATES

While experimental results with proton-proton collisions [11] are compatible with having achieved perfect overlap between the electron and proton beams, possible deviations are difficult to estimate from these measurements. We explore in this section to what extent maximizing the eBSD counting rates may result in imperfect overlap. We identify two sources of systematic errors, one in the horizontal and one in the vertical alignment, and we estimate the magnitude of these errors with simulations for the specific example of the $^3$He tests for which we have the necessary data. Table 1 lists the relevant parameters.

Table 1- Beam parameters during the $^3$He tests

| Location | | Interaction region | eBSD detector |
|---|---|---|---|
| Magnetic field | (T) | 4.00 | 0.25 |
| e-beam rms size | (mm) | 0.375 | 1.50 |
| $^3$He-beam vertical rms size | (mm) | 0.32 | |
| $^3$He-beam horizontal rms size | (mm) | 0.46 | |

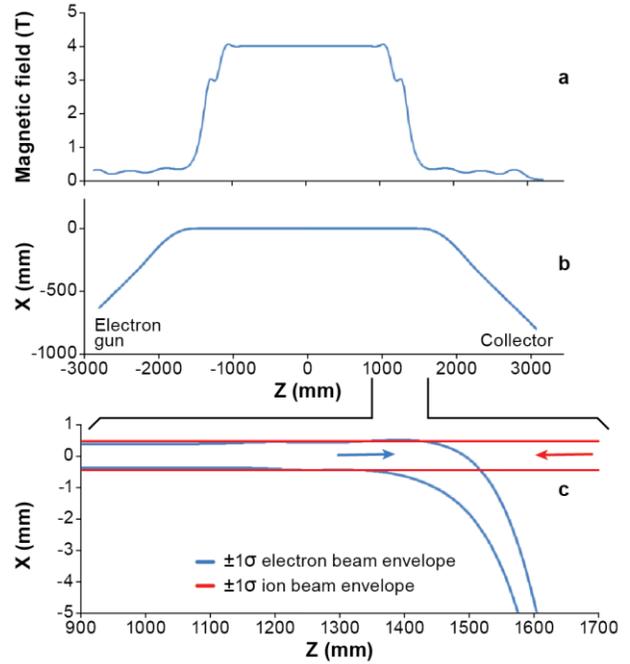

FIG.18 Magnetic field profile (a) and electron beam trajectory (b) starting at the electron gun and ending at the electron collector. In the magnified view (c) we show the ±1 sigma electron beam and ion beam profiles in the region where the electron beam starts curving away on its way to the collector.

For the horizontal alignment, there is an obvious bias due to the fact that at the entrance and exit of the electron-ion beam overlap region, the electron beam deviates from the straight trajectory on its way from the electron gun and to the electron collector. As shown in Fig.18 there are regions at both ends of the overlap region where the electron beam becomes larger and curves away from the ion beam trajectory. Backscattered electrons from these regions will make an asymmetric contribution to the counting-rate curve when scanning one beam across the other.

In Fig. 18 c we show a magnified view of the area where this asymmetry arises at the electron collector side of the interaction region. The electron beam transport is symmetric around the center of the solenoid since the magnetic field is symmetric (see Fig. 18 a. In Fig 18 c we also show a ±1 sigma ion beam profile centered with respect to the electron beam in the solenoid. We estimate the relative counting rate variation as function of offset by computing the convolutions of these two beams (assumed to be Gaussian) as function of offset in 0.1 mm steps. The result is shown in Fig. 19, and compared to the simulated counting rate profile in the absence of the asymmetric contribution.

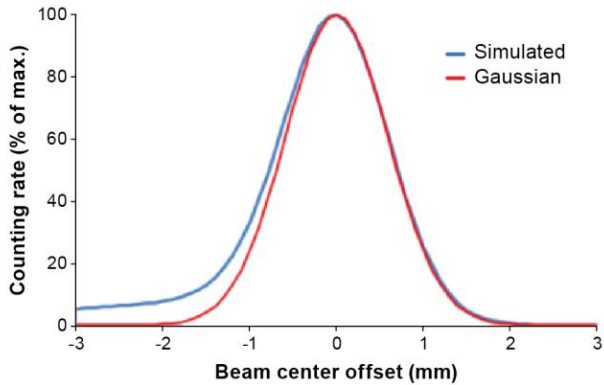

FIG.19 Simulated relative counting rates as function of horizontal electron-ion beam offset for the case illustrated in Fig. 18. Also shown (in red) is the ideal convolution of the two Gaussian beams in absence of the asymmetric contributions.

The counting rate asymmetry as function of horizontal beam offset is clearly visible. The peak position was calculated with a quadratic fit over a ±0.2 mm range, following the procedure used by the automatic adjustment software [29]. The peak offset is 0.018 mm for this example where the rms widths of the electron and ion beams are 0.375 mm and 0.46 mm respectively. This shift is not expected to have any measurable consequences upon the e-lens performance in this case. If much larger deviation should occur in other situations, corrections could be computed and applied.

In the vertical direction, there is a bias introduced by the detection efficiency dependence on the vertical position of the Coulomb interaction point. In other words, a backscattered electron originating from a point located at some distance below the axis common to the electron and ion beams, will have a slightly different detection probability compared to another electron originating at a point located above that axis. This effect could be measured by vertically displacing the interaction region, i.e. both the electron and the ion beams together, and monitoring associated counting-rate changes. While this approach may be attempted in the future, we will here arrive at an estimate based on the measured counting-rate dependence on the vertical position of the detector shown in Fig. 12. The slope of this curve at the operating position at 24 mm is ~6.8 % per mm. This detection efficiency slope at the detector location can be converted to an equivalent efficiency slope in the interaction region. To that effect, we take into account the adiabatic invariance of the flux through the electron orbits [26] which leads to similar projections of the electron orbits onto planes perpendicular to the field at the detector position and in the interaction region. These similar projections differ by a scale factor equal to the ratio of the square roots of the magnetic field strengths at these two locations. In the present case that scale factor is 4.0 (see Table 1). That in turn means that the slope of the detection efficiency translated to the interaction region will be 6.8 ×4 =27.2 %/mm.

The counting rate as function of $^3$He beam position will be the usual convolution of two Gaussians, but modified here by the efficiency which we approximate as a linear function of the position with the slope 27.2%/mm obtained above. Using this value for the parameter k, and the vertical rms beam sizes $\sigma_{He}$ and $\sigma_e$ from Table 1, we use Eq. 2 to calculate values proportional to the eBSD counting rate as function of the He beam position:

$$N(y_{He}) = \int_{-\infty}^{\infty} \exp\left[-\frac{y_e^2}{2\sigma_e^2}\right] \times \exp\left[-\frac{(y_e - y_{He})^2}{2\sigma_{He}^2}\right]$$

$$\times (1 + k\, y_e)\, dy_e \qquad (2)$$

The approximation that was made in this estimate is neglecting the small variations in vertical drift corresponding to variations in longitudinal electron velocity. These variations are small because of electron momentum conservation and because of the small angles between the electron trajectories and the magnetic field lines in the region of the bend. For the example shown in Fig. 4, these angles are around ~8⁰. The result is an approximately Gaussian curve of rms width $\sigma \sim \sqrt{\sigma_e^2 + \sigma_{He}^2}$, with its maximum displaced by 0.028 mm. This is less than 10% of the rms widths of either beam. A correction is not necessary in this case.

In other situations, in particular for beams of much larger widths, corrections could be computed or measured as outlined above, and applied by introducing a position correction after maximizing the counting rate.

## VII. FUTURE POSSIBILITIES

In this section, we briefly present a few preliminary ideas on how the detection of energetic scattered electrons could be used in other beam diagnostic applications. These ideas are based on the previous extensive use of electron beams as diagnostic tools, documented in the literature, and on the results and experience gained during the design, implementation and application of this new approach.

### A) eBSDs used with hollow electron beams as possible halo monitors and beam alignment tools.

Hollow electron beams have been tested as collimators or halo collimators in the Tevatron [34, 35, 36, 37] and are being considered as an option to complement the LHC collimation system [13]. Here, we suggest that the backscattered electrons from the proton electron collisions could be detected and used for halo diagnostics and for centering the proton beam [38].

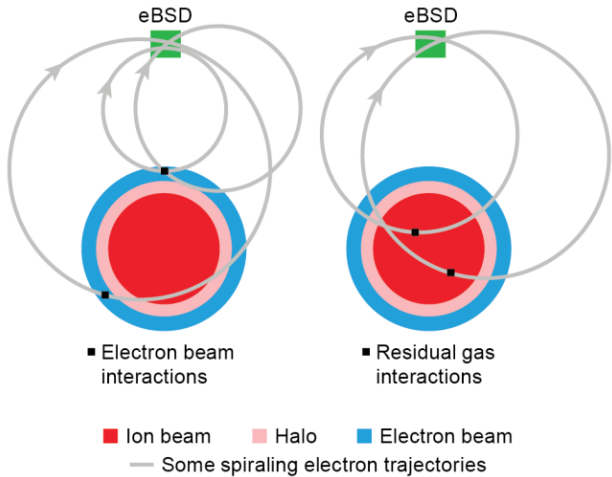

FIG. 20 Schematic illustration of the potential use of eBSDs for monitoring the halo of an intense ion beam. The beam propagates in a direction perpendicular to the plane of the figure. The electrons generated by the interaction of the ions with the hollow electron beam spiral along the magnetic field towards the detector generating a counting rate approximately proportional to the overlapped halo intensity (left). The ion beam interacting with the residual gas produces a background (right). Excellent vacuum is required to minimize this background (see text).

The arrangement would be similar to the BNL electron lenses, but additional thought is required to determine the best way to merge the beams and to separate them after the interaction region without producing unduly large background counting rates in the detectors.

A schematic illustration of the interaction between the halo protons with the electrons in the hollow beam is shown in Fig. 20. In reality, two or four equidistant detectors surrounding the beam would probably be used. The core of the proton beam will also produce energetic electrons by collision with the atomic electrons of the residual gas. This is the principal source of background and will determine the ultimate sensitivity for halo detection. For a rough estimate of this background we note that a 4 keV electron current density of 1 A/mm$^2$ has an electron density equal to the electron density in $2.15\times10^{-6}$ Torr of $H_2$ at room temperature. For an example of a round Gaussian ion beam of rms width σ we conclude that for a hollow, 1 A/mm$^2$ electron beam extending from 4σ to 5σ the signal-to-background ratio would be approximately 1 if the vacuum is $7\times10^{-10}$ Torr at room temperature. A better vacuum and/or a more intense electron beam will improve this signal-to-background ratio. Exceptionally good vacuum in a room-temperature chamber should be achievable in a beam pipe section pumped by a continuous longitudinal cryo-pumped antechamber as mentioned for example in reference [39]. If that is impractical, a Non Evaporable Getter (NEG) coated and activated beam pipe would be excellent too. The use of a warm chamber with an adjacent distributed cryo pump is appealing since the quantity of interest to reduce the background is the gas density which, at constant pressure, is inversely proportional to the absolute temperature.

A technique that can be used to extend the dynamic range of these measurements involves modulating or pulsing the electron beam. Depending on counting statistics, results could be obtained with signal to background ratios as small as a few percent.

Figure 21 shows schematically the topology of three possible implementations. The first one seems elegant and appealing but it may be difficult to implement an annular cathode surrounding the proton beam. The second one has the same geometry as the existing electron lenses, but the ion beam intersects the electron beam on the collector side producing unwanted background. Finally, the third option solves these problems by locating the gun with the annular cathode to one side, and uses an annular collector surrounding the proton beam which should be possible to implement. This appears to be a viable option for a system that could be used as a halo monitor and as a beam alignment tool.

Aligning electron "wires" proposed as LHC long-range beam-beam compensators [40] may be achieved in a similar way, and without the complication of annular cathodes and collectors. The beams would be aligned by first overlapping them and then separating them by a known distance. If the electron beam remains partly in the halo of the proton beam, continuous monitoring would also become possible.

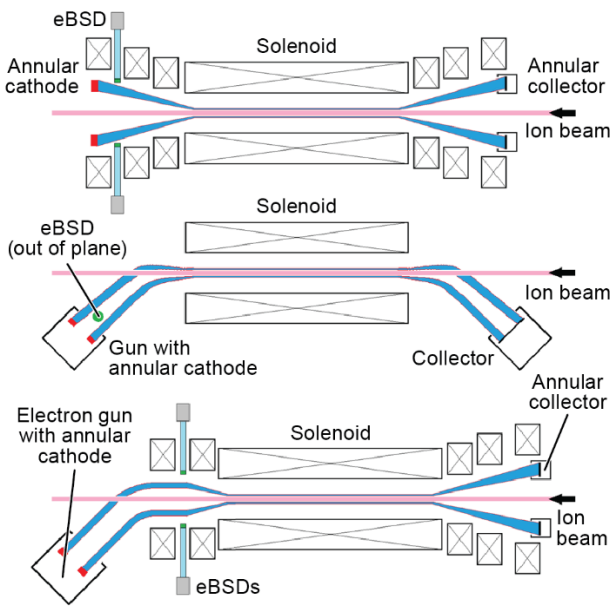

FIG. 21 Three possible configurations for using eBSDs for halo monitoring and beam alignment of hollow electron beam systems. The central long solenoid is the same strong field superconducting solenoid in each case, similar to the 6T ones used in the Brookhaven RHIC electron lenses. The smaller and weaker room-temperature solenoids, indicated schematically by the small rectangles, guide the electron beam from the gun to the central solenoid and from there to the collector (see Fig. 3). Some of these guiding solenoids have been omitted for clarity. Only two eBSDs are shown in the bottom configuration, but there could be four at 90º intervals for continuous halo monitoring and beam centering. The bottom configuration seems to be the most feasible one (see text).

## B) Concept of a Coulomb Scattering Electron Wire (CSeW) beam profile monitor.

Electron beams that are not collinear with the relativistic ion beam will also generate energetic scattered electrons that can in principle be used for beam diagnostics. An example is schematically shown in Fig. 22. A ribbon shaped electron beam propagates at a right angle to the ion beam guided by a weak magnetic field (2B) that affects the ion beam only slightly. This slight perturbation is compensated by the field B generated by the other two split solenoids.

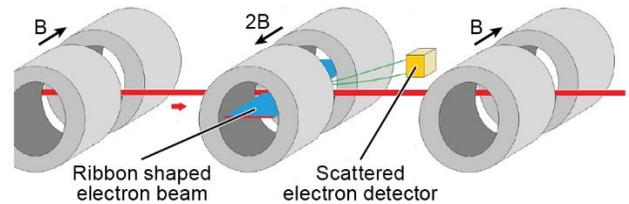

FIG.22 Schematic illustration of a Coulomb Scattering Electron Wire beam profile monitor (see text).

The trajectories of the scattered electrons are bent in the field of the central split solenoid and some of them reach a scintillation detector through a vacuum window (not shown in the picture). Maximum intensity corresponds to optimal overlap. The beam profile can be explored by stepwise deflections of either the electron or the ion beam.

In contrast to conventional electron wire profile monitors [6, 7, 8], the profile is determined here by measuring the counting rates of the scattered electrons and not by detecting the deflection of the electron trajectories, largely suppressed here by the transverse magnetic field. Potential advantages are that the measured profiles are largely independent of the beam intensity and that profiles are obtained directly as deflection-dependent counting rates. For relatively long bunches, the arrival time of the scattered electrons can be used to measure the time structure and head-tail position differences for each bunch. Two such systems, one horizontal and one vertical, would provide a rather complete characterization of the bunch through non-destructive measurements.

### C) Electrons scattered from residual gas atoms for beam diagnostics.

The interaction of particle beams with residual gas atoms and molecules is often used for measuring beam profiles such as in ionization profile monitors (see e.g. [41]) and fluorescence profile monitors (see e.g. [42]). Recently, a beam-gas vertexing technique [43] was used to characterize LHC beam properties by high precision tracking of particles from nuclear interaction with a small amount of gas injected into the vacuum chamber [44, 45].

We suggest here that detecting energetic scattered electrons is another good way to exploit the interaction with residual gas for beam diagnostic purposes. The cross sections for Coulomb interactions are orders of magnitude larger than for nuclear cross sections. Much less gas will therefore be required. As an example, we show a conceptual design for a beam position monitor for eRHIC [46], the proposed BNL ERL-based electron-ion collider. This is only one of several possibilities for the difficult task of monitoring the position of up to 24 side-by-side electron beams circulating in the same vacuum chamber and separated in time by as little as 2 ns. As shown schematically in Figure 23, two fast, position-sensitive channel-plate detectors detect the scattered electrons through sets of parallel plate collimators which are necessary to define the plane of the trajectories. Thin foils in front of the detectors stop low energy electrons from generating spurious signals. A second set of detectors and collimators, at right angles to the first one, could, in principle, be located in the same chamber. The detection by the channel plates is fast and the position resolution will be defined by the acceptance angle of the collimators.

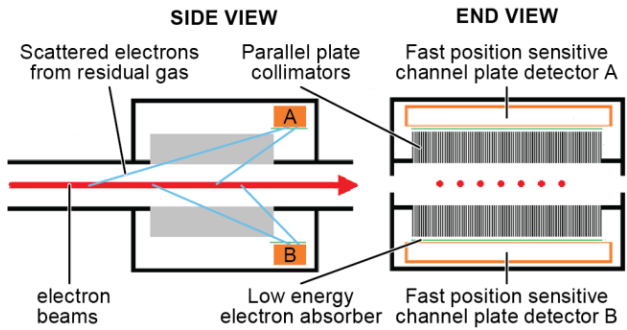

FIG. 23 Concept of one of the possible approaches to detect several side-by-side orbits in one of the fixed-field alternating gradient (FFAG) arcs of the electron-ion collider, eRHIC, which will be proposed as a successor to RHIC. Fast, position-sensitive channel-plate detectors respond to energetic electrons collimated by parallel plate collimators (see text).

## VIII. CONCLUSIONS

Most instruments used for particle accelerator beam diagnostics are of the analog type where often small signals are transmitted through long cables, amplified and digitized. The few instances when particle detecting and counting techniques can be used, offer the advantages of greater dynamic range and greater noise immunity which is particularly important in the harsh environment of high energy particle accelerators.

The detection of energetic electrons generated through Coulomb scattering by relativistic ions offers new possibilities for relativistic ion beam diagnostics. The fact that these electrons can traverse vacuum windows with relatively minor energy loss allows the convenient use of simple detectors such as scintillation detectors that are cumbersome to use in vacuum. The easy replacement of the detectors without disturbing the vacuum is also an important advantage.

We have shown here the successful application of such a system, used for the alignment of electron and ion beams in the RHIC electron lenses at BNL. A counting rate dynamic range of about five orders of magnitude has been utilized so far. Given the fast response of the utilized scintillators and counting electronics, larger dynamic ranges are available. A likely improvement for future systems of this type will be the utilization of silicon photomultipliers which are not sensitive to magnetic fields. Shorter light guides and less magnetic shielding should simplify the design and improve the performance.

We also outlined ideas for other possible beam diagnostic applications based on energetic electrons produced by Coulomb scattering by relativistic ions.

## ACKNOWLEDGEMENTS

We are grateful to the RHIC operations crew for their expert assistance in the development, testing and implementation of this new system. We would like to thank the referees for many valuable suggestions. In particular, the idea of modulating the electron beam, described in section VII A could lead to an important sensitivity improvement and the possibility mentioned in section VI of displacing both beams simultaneously to better evaluate a systematic error may be implemented in the future. This work was supported by Brookhaven Science Associates, LLC, under Contract No. DE-AC02-98CH10886 with the U.S. Department of Energy.